**Dayside magnetospheric and ionospheric responses to a foreshock transient on June 25, 2008: 2. 2-D evolution based on dayside auroral imaging**

Boyi Wang[1,2], Yukitoshi Nishimura[3,1], Heli Hietala[4], Xiao-Chen Shen[5,6], Quanqi Shi[5], Hui Zhang[7], Larry Lyons[1], Ying Zou[2,8], Vassilis Angelopoulos[4], Yusuke Ebihara[9], Allan Weatherwax[10]

**Affiliations:**

1. Department of Atmospheric and Oceanic Sciences, University of California, Los Angeles, California, USA

2. Department of Astronomy and Center for Space Sciences, Boston University, Boston, Massachusetts, USA

3. Department of Electrical and Computer Engineering and Center for Space Sciences, Boston University, Boston, Massachusetts, USA

4. Department of Earth, Planetary and Space Sciences, University of California, Los Angeles, California, USA

5. Optical Astronomy and Solar-Terrestrial Environment, School of Space Science and Physics, Shandong University, Weihai, China

6. High Altitude Observatory, National Center for Atmospheric Research, Boulder, Colorado, USA

7. Geophysical Institute, University of Alaska Fairbanks, Fairbanks, Alaska, USA

8. Cooperative Programs for the Advancement of Earth System Science, University Corporation for Atmospheric Research, Boulder, Colorado, USA

9. Research Institute for Sustainable Humanosphere, Kyoto University, Kyoto, Japan

10. School of Science and Engineering, Merrimack College, North Andover, Massachusetts, USA

**Corresponding author:** Boyi Wang (bywang@ucla.edu)

**Key Points:**

- We for the first time show auroral evolution due to a foreshock transient in high-resolution 2D imaging.
- By mapping to the magnetosphere, the imaging was able to determine the size (a few RE in Y) and propagation (duskward).
- Equivalent currents show a pair of FACs. Its duskward propagation is consistent with aurora. The response resembles to sudden commencements.

**Abstract**

The foreshock region involves localized and transient structures such as foreshock cavities and hot flow anomalies due to solar wind-bow shock interactions, and foreshock transients have been shown to lead to magnetospheric and ionospheric responses. In this paper, the interaction between a foreshock transient and the magnetosphere-ionosphere system is investigated using dayside aurora imagers revealing structures and propagation in greater detail than previously possible. A foreshock transient was detected by THEMIS-B and C during 1535-1545 UT on June 25, 2008. THEMIS-A, D and E observed magnetopause compression, cold plasma enhancement and ULF waves in the dayside magnetosphere. The all-sky imager (ASI) at

South Pole observed that both diffuse and discrete aurora brightened locally soon after the appearance of this foreshock transient. The diffuse aurora brightening, which corresponded to a region a few Re size in GSM-Y in the equatorial plane, propagated duskward with an average speed of ~100 km/s. Soon after the diffuse aurora brightened, discrete aurora also brightened and extended duskward, which was consistent with the motion of the foreshock transient as it swept through the magnetosheath while impacting the magnetopause. Equivalent horizontal currents measured by magnetometers revealed a pair of field-aligned currents (FACs) moving duskward consistent with motion of the discrete aurora patterns. We conclude that the high-resolution and two-dimensional observation of auroral responses by ground-based ASI can help to estimate the evolution and propagation of upstream foreshock transients and their substantial impacts on the magnetosphere-ionosphere coupling system, including magnetospheric compression and currents in the ionosphere.

**1 Introduction**

Earth's foreshock is a dynamic plasma interaction region between the solar wind and reflecting particles from the bow shock, forming under quasi-parallel shock conditions, when the Mach number of the solar wind is high and the IMF Bx dominates [*Omidi et al.*, 2009]. Reflected super-thermal ions interacting with upstream solar wind particles trigger instabilities and waves [*Eastwood et al.*, 2005; *Paschmann et al.*, 1981; *Fairfield et al.*, 1990]. These transient kinetic processes in the foreshock can significantly modify the solar wind right in front of the bow shock and form transient structures in the magnetosheath that can potentially influence the magnetosphere.

Transient kinetic processes in the foreshock lead to hot flow anomalies (HFAs), spontaneous hot flow anomalies (SHFAs), foreshock bubbles (FBs), foreshock cavities, foreshock cavitons, foreshock compressional boundary, density holes, and short, large-amplitude magnetic structures (SLAMs). Most of these transients are associated with depletions in the density and magnetic field strength and compressions at edges [*Schwartz, 1995*; *Sibeck et al.*, 2002; *Blanco-Cano et al.*, 2011; *Turner et al.*, 2013]. HFAs, FBs, foreshock cavities and density holes are often associated with IMF discontinuities in the solar wind [*Lin*, 1997; *Omidi et al.*, 2010, 2013; *Wang et al.*, 2013; *Liu et al.*, 2015], while other transients could evolve from ULF waves or other physical processes in the foreshock [*Le et al.*, 1992; *Schwartz et al.*, 1992].

Unlike large-scale solar wind changes such as interplanetary shocks, foreshock transients are localized and short-lived. Their size and lifetime are between an ion gyroradius and a few RE, and from a few seconds to a few minutes, respectively [*Parks et al.*, 2006; Omidi et al., 2009, 2010; *Billingham et al.*, 2011]. HFAs are associated with flow deflections in the magnetosheath and deformation of the magnetopause, which results in various impacts on the magnetosphere including plasma injection into the cusp generation of field-aligned currents (FACs), triggering of global-size Pc3 waves in the magnetosphere, and creating traveling convection vortices (TCVs) in the ionosphere [*Sibeck et al.*, 1999; *Jacobsen et al.*, 2009; *Omidi et al.*, 2010; *Zhao et al.*, 2017]. FBs also lead to transient magnetopause compression and ULF waves in the magnetosphere [*Omidi et al.*, 2010; *Hartinger et al.*, 2013; Archer et al., 2012].

Although the studies above showed that foreshock transients are geoeffective, it has been difficult to accurately detect structure and evolution of the influences on the magnetosphere-ionosphere (MI) coupling system, because measurements are often limited to several data points or by global imaging with coarse resolution.

On the other hand, ground-based imaging can provide high-resolution measurements of discrete and diffuse aurora in response to dayside solar wind disturbances, such as poleward moving aurora forms (PMAFs) as the response of the flux transfer events (FTEs) and shock aurora [*Tsurutani et al.*, 2001; *Vorobjev et al.*, 2001; *Sandholt et al.*, 2003; *Zhou et al.*, 2003, 2009; *Mende et al.*, 2009; *Holmes et al.,* 2014; *Wang et al.*, 2016;]. Such auroral signatures can reveal location, size and propagation of dayside disturbances in much greater detail than by other means of measurements. Discrete aurora is generated by electrons accelerated along magnetic field lines, and is associated with upward FACs. Thus discrete aurora can be used to highlight upward FACs associated with dayside magnetopause disturbances [*Maltsev et al.*, 1975; *Cowley*, 2000]. Diffuse aurora is generated by electron cyclotron harmonic (ECH) waves and whistler-mode waves scattering energetic electrons into the loss cone [*Ashour-Abdalla* and *Kennel*, 1978; *Devine and Chapman*, 1995; *Horne et al.*, 2003; *Ni et al.*, 2008; *Li et al.*, 2009, 2011; *Thorne et al.*, 2010; *Nishimura et al.*, 2010, 2013; *Tao et al.*, 2014; *An et al.*, 2017]. Thus diffuse auroral brightenings can display the area of the dayside magnetosphere where particle anisotropies are changed by magnetosphere compression, thereby leading to enhanced wave power and subsequent electron scattering.

Similarly, since foreshock transients also induce magnetopause disturbances, auroral brightenings are expected to occur. *Sibeck et al.*, [1999] presented aurora brightening followed an HFA. However, their study was not able to identify the type of aurora or its evolution (which was rapid and thus time-aliased) due to limited spatial and temporal resolution of space-based imaging.

The availability of multi-point measurements near the magnetopause from the THEMIS satellites together with high-resolution imaging (spatial and temporal) of the dayside aurora enables far more detailed studies of these phenomena than ever before. In the present paper, we demonstrate the power of this method and reveal some of the auroral characteristics associated with the interaction of foreshock transients with the magnetosphere. We present a foreshock transient event that occurred during a conjunction between the THEMIS satellites and an all-sky imager (ASI) at South Pole on the dayside. This conjunction event provides a simultaneous multi-point observation from interplanetary space to aurora regions through the foreshock, magnetosheath, and magnetosphere. 2-D imaging gives us the information of structure and evolution of both diffuse and discrete aurora patterns in its field of view (FOV). By mapping aurora patterns to the equatorial plane in the magnetosphere, the mapped pattern can reveal the size and evolution of the related upstream activities in a 2-D perspective. *Shen et al.* [2017, "Dayside magnetosphere and ionosphere response to a foreshock transient: 1. FLR observed by satellite and ground-based magnetometers", submitted to JGR, hereinafter Paper 1] investigates properties of ULF waves driven by the foreshock transient, while the present study focuses on 2-D evolution deduced from optical observations.

**2 Instruments and event selection**

The THEMIS satellites measure plasma and fields in the solar wind, near the magnetopause and in the dayside magnetosphere. Magnetic field, velocity and frequency of magnetic field can be obtained by Flux Gate Magnetometer (FGM), Electrostatic Analyzer (ESA)/ Solid State Telescope (SST) and Search Coil Magnetometers (SCM) respectively. The satellite observations are used to observe the upstream activities and their corresponding responses in the magnetosphere.

This study also uses a monochromatic ASI at South Pole station in Antarctica [*Ebihara et al.*, 2007] to identify dayside discrete and diffuse aurora. As shown by the mapping to equatorial plane through the T01 model in Figure 1C, the imager covered an area that swept across noon during in this event. This ASI records 557.7 and 630.0 nm wavelength images every ~40 seconds. Red-line (630.0 nm) emissions are sensitive to soft (<1 keV) electron precipitation and can be used to observe discrete aurora commonly seen near the dayside open-closed field line boundary [*Lorentzen et al.*, 1996]. Red-line data are mapped to 230 km altitude. This is a representative altitude of emission due to low-energy precipitation at ~150-300 km [*Solomon et al.*, 1988]. Green line (557.7 nm) emissions equatorward of red-line emissions can be used to identify diffuse aurora, which comes from high-energy (> 1 keV) precipitation on closed field lines [*Meng et al.*, 1979; *Lorentzen et al.*, 1996; *Lorentzen and Moen*, 2000; *Sandholt et al.*, 2002]. Green-line data are mapped to 110 km altitude. This is a representative altitude of emission due to high-energy precipitation [*Mende et al.*, 1993].

To quantify a characteristic size and speed of auroral structures, we identify an area above 1/e of the maximum luminosity above the background for each snapshot in each wavelength. Here, the background is defined as a median luminosity at each pixel within $\pm 10$ minutes from each observation time. Since the auroral brightening of interest lasted less than 10 minutes, this background gives a representative value of luminosity before/after the event at each pixel. The choice of 1/e threshold is made because latitudinal and longitudinal profiles of emission often show roughly Gaussian-like distributions.

We also utilize the ground-based magnetometers at high latitudes in the northern hemisphere to calculate the horizontal currents and geosynchronous GOES and interplanetary

ACE spacecraft measurements. The ground magnetometer data is from SuperMAG at 1-min time resolution.

**3 THEMIS satellite observations**

**3.1 A foreshock transient**

Figure 1 summarizes THEMIS satellite observations and also shows red and green line imager keograms. Figures 1Ad and 1Ah show that during 1536-1541 UT and 1537-1545 UT, a foreshock transient, which appears to propagate towards the bow shock with suprathermal particles, is identified to cross THEMIS-B and C, respectively. Both satellites observed two discontinuities and suprathermal ions within them. The first discontinuity was observed by THEMIS-B and C at 1534:40 and 1536:58 UT, respectively. The second discontinuity was observed by THEMIS-B and C at 1540:45 and 1544:18 UT, respectively.

To identify the normal of the discontinuities, the timing method based on ACE, THEMIS B and C satellites is utilized. The first discontinuity was observed by ACE at 1420:48 UT. Based on those 3 satellites, the normal of this discontinuity is about (0.37, -0.93, -0.06) in GSM coordinates. It would arrive at the dawnside first and sweep duskward. The second discontinuity was observed by ACE at 1425:06 UT. The calculation shows that the normal of this discontinuity is about (0.37, -0.87, -0.34) in GSM. It would also arrive at the dawnside first and sweep duskward. The second and third columns of Table 1 show the timings of arrival of these 2 discontinuities at the locations of 5 satellites. The magnetic fields in the normal direction of those 2 discontinuities were less than ~0.1 of the total magnetic field, and their normals calculated from the cross-product method [*Hudson*, 1970] were consistent with those from the timing method. This indicates that these 2 discontinuities were tangential discontinuities.

The cone angles of the magnetic field within the two tangential discontinuities at THEMIS-B and C are ~45° and ~26°, respectively, which are relatively lower than in the background solar wind (~60°). The lower cone angles indicate that the magnetic field within these two discontinuities were connected to the bow shock and therefore the ions are reflected and accelerated, forming the foreshock region within the discontinuities. In addition, the depression of magnetic field and the density, small compressions at the edges but no significant flow deflection is found in both satellite data. Those signatures indicate that this foreshock transient is probably a foreshock cavity. In addition, during the foreshock observations, the magnetic field became more fluctuating than the background (Figures 1Ab and 1Af). The foreshock region was also associated with two density reductions sandwiched by density increases (Figure 1Aa). THEMIS-C observed essentially the same structures with a few minute propagation delay, except that the amplitudes were larger and the disturbance lasted longer. Figure 1Af presents a significant Bz reduction (~1537 UT) followed by an increase in Bz (~1545 UT), associated with low-frequency fluctuation. Although velocity changes were small, the foreshock region was associated with small decreases in |Vx| (Figure 1Ag). Those transient foreshock signatures are consistent with a proto-HFA by *Zhang et al.*, [2010] though the velocity reduction is small. The small velocity change at the satellite location could be because the bulk of the structure may be localized near the bow shock and the satellites only detected an early stage or a portion of the structure.

At the same time, there was no appreciable change in the pristine solar wind density and velocity before and after the foreshock transient, and the dynamic pressure in the ACE data during this period was almost constant. Thus, it cannot be an interplanetary shock, so that this foreshock transient is probably a foreshock cavity or possibly an HFA, which was sandwiched by two tangential discontinuities.

### 3.2 Magnetosheath and Magnetospheric signatures

During the same period, THEMIS-A was located at ~15.5 MLT and ~9.5 Re away from the earth in the magnetosphere. Figures 1Ba-Bd show THEMIS-A observation of the density, detrended magnetic field, flow velocity and ion energy flux, respectively. Here, Figure 1Bb shows the magnetic field that is detrended by the T01 model. During 1544-1546 UT, high energy (>1 keV) particles almost disappeared while low-energy particle fluxes (<1 keV) became higher, and the magnetic field decreased. Those signatures indicate a magnetopause crossing and THEMIS-A was in the magnetosheath during that 2 minutes. Thus, at the position of THEMIS-A, there was a brief magnetopause compression at ~1544 UT. Additionally, the GSM-x component of velocity in the magnetosheath approached to ~100 km/s, which was fast compared with the background flow (~30-40 km/s in this case). This magnetosheath fast flow was identified as a magnetosheath high speed jet (HSJ) by *Dmitriev and Suvorova* [2012]. It could be due to the impact of the foreshock transient [*Archer et al.*, 2014] and also can cause magnetospheric compression and fast channels in ionosphere [*Hietala et al.*, 2012; *Archer et al.*, 2013].

The observation by THEMIS-D, which was located at ~14.0 MLT and ~7.5 Re away from the earth in the magnetosphere are shown in Figures 1Be-Bh with the same format as Figures 1Ba-Bd. Starting from ~1540 UT, THEMIS-D observed a gradual increase in the GSM-Z component of detrended magnetic field followed by ULF waves that are observed in both magnetic field and velocity. Those two signatures, which are generally considered to be the responses to magnetopause compressions [*Takahashi et al.*, 1988; *Lysak et al.*, 1992], and also occurred soon after the occurrence of the observed foreshock transient. Paper 1 shows that the compressional waves induced oscillations perpendicular to the magnetic field and that those are a field line resonance (FLR). The black line in Figure 1Bh shows the kinetic energy of the bulk flow moment and demonstrates that the energy variation apparent in the spectra is due to the flow velocity oscillations caused by the impact of the transient.

Other satellites, i.e., GOES-12, 10 and THEMIS-E (not shown), located at ~10.3, 11.3 and 13.2 MLT, respectively, also observed compressional signals in magnetic field (Bz increase and ULF waves) at ~1533, 1534 and 1539 UT, respectively. The timings when all 5 satellites (GOES-12, 10, THEMIS-E, D and A) observed magnetospheric compressional signatures are listed in the fourth column of Table 1. Both the foreshock transient and the magnetospheric signatures took 11-14 minutes to propagate from 10.3 to 15.5 MLT. This indicates that the magnetospheric compression is most likely driven by the foreshock transient, and that the magnetosphere took ~10 minutes to respond to the foreshock transient.

**4 Evolution of diffuse aurora brightening**

Figures 1Da and 1Db show east-west (EW) keograms of red-line and green-line emission, respectively from the South Pole imager data (at -74° MLAT). The green-line emission shows that diffuse aurora began to brighten at ~1537 UT and started to propagate duskward. At ~1542 UT, the red-line emission shows that the discrete aurora began to brighten and also propagated duskward, a pattern similar to that of the diffuse aurora.

Figure 2 shows snapshots of background-subtracted green-line emission observed by the ASI at South Pole station in magnetic coordinates. The yellow contour lines in Figure 2 present the characteristic boundary of emission using the method described in section 2. The FOV was dark until 1536:14 UT. A bright pattern showed up at the western edge of the FOV at 1538:23 UT. The position of this initial pattern is at ~4° MLON and ~-73° MLAT, at ~10 MLT in the ionosphere. The following snapshots show that the structure became larger and brighter until 1541:59 UT. The pattern extended in size to at most ~25° in MLON and ~5° MLAT in width. Those correspond to ~900 and ~590 km in the east-west and north-south directions, respectively, at the altitude of ~110 km in the ionosphere. Then, the brightening pattern started to become dimmer, smaller and disappeared near the eastern edge of the FOV at 1545:35 UT. During the entire process, this diffuse aurora structure propagated duskward by ~3 h MLT.

The initiation of the diffuse aurora occurred 2 minutes after the foreshock transient measurement by THEMIS-C. Although this time lag is somewhat short compared to a signal propagation from the satellite to the ionosphere, as shown later, it should be noted that the auroral signature was detected first at pre-noon and propagated to post-noon, while the satellite was at post-noon. This propagation indicates that the satellite did not detect the earliest signature at pre-noon but encountered the signal later when the structure reached post-noon. This is also consistent with the THEMIS-A observation that the magnetopause crossing at post-noon was later than the initiation of aurora at pre-noon and that the plasma flow at the magnetopause was predominantly duskward.

The diffuse auroral structures shown in Figures 2b-k are mapped to the equatorial plane by the T01 model [*Tsyganenko*, 2002]. The red star in each panel of Figure 3 shows a mapped position of the centroid of the diffuse aurora structures seen in the ionosphere. The mapped patterns show a more significant expansion in the GSM-Y direction than in the GSM-X direction: The pattern was initially centered at ~[8.2, -2.7] Re in GSM with a width of 1.4 Re in the X direction and 2.2 Re in the Y direction (Figure 3a). Figures 3b-j show that the mapped pattern expanded to at most 4.3 Re in GSM-Y at 1540:33 UT and then decayed to 0.8 Re in GSM-Y at 1544:52 UT. The range of 0.8-4.3 Re is overall comparable to the azimuthal size of foreshock transients (a few Re) [*Lin et al.*, 2002; *Parks et al.*, 2006; *Zhang et al.*, 2010; *Turner et al.*, 2013; *Plaschke et al.*, 2016]. The mapped pattern also shows a duskward propagation to ~[8.2, 2.7] Re in GSM.

Figures 4a-4c plot the magnetic longitudes, GSM-Y position and MLT, respectively, of the centroids of the mapped diffuse aurora patterns (shown as the larger blue squares). The smaller squares show the corresponding maximum and minimum values of each mapped patterns, corresponding to their leading and trailing edges, respectively. We also did linear fitting on each sets of data (shown as magenta and orange lines). Since the diffuse auroral structure in the first two (in Figures 2b and 2c) and the last two (in Figures 2j and 2k) snapshots were close to the edge of the FOV, the linear fitting of centroids was done on the six points from 1539:50-1943:25 UT. Similarly, since the leading edges of those patterns are only shown in the first eight snapshots (Figures 2b-2i), the linear fitting of the leading edges was done on the first eight points. And since the trailing edges of those patterns are only shown in the last eight snapshots (Figures 2d-2k), the linear fitting of the trailing edges was done on the last eight points.

From Figure 4a, the average velocity in the ionosphere that is obtained by the linear fitting is ~7.1°/min or ~235 km/min in the ionosphere. The speeds of the leading and trailing edges are ~6.0 and ~7.9°/min, respectively, or ~198 and ~261 km/min, respectively, in the ionosphere. Figure 4b shows that the average velocity of the mapped patterns in GSM-Y direction in the equatorial plane is ~117.4 km/s. The speeds of their leading and trailing edges are ~100.5 and ~144.7 km/s, respectively. Figure 4c shows that the average velocity of MLTs of the mapped patterns in the equatorial plane is ~0.50 h/min. The speeds of their leading and trailing edges are ~0.43 and ~0.65 h/min, respectively.

While the numbers mentioned here depends on mapping altitudes, the structures measured near zenith (~1540-42 UT) should be most reliable. The size and propagation speed are nearly constant except in the first and last couple of snapshots as shown in Figure 4 where edge effects are significant. Thus the numbers we obtained reasonably represent the ionospheric size and speed of the diffuse aurora induced by the foreshock transient.

Based on the results shown above, we obtain the timings when the diffuse aurora brightening arrived at the MLTs of the 5 satellites and listed them in the last column of Table 1. Similar to the foreshock transient and the magnetospheric signatures, diffuse aurora also took ~12 minutes to propagate from 10.3 to 15.5 MLT. This indicates that the diffuse aurora traces the upstream motion well. It is also noted that the diffuse aurora took ~3-4 minutes to respond to the magnetospheric signatures. This time delay is reasonable to include the response time of kinetic processes in magnetosheath, wave-particle interaction to the magnetosphere compression, M-I communication time, and the imager time resolution (~44s).

## 5 Evolution of discrete aurora brightening

Figure 5 shows snapshots of red-line emission data observed by the ASI at South Pole station. Much of the emissions is seen a few degrees poleward of the diffuse aurora. As described in Section 2, red-line emission mainly sees discrete aurora due to low energy precipitation near the open-closed boundary. Faint emissions are also seen in the region of the green-line diffuse aurora and propagate duskward (Figures 5a-f). Those are likely due to low-energy portion of precipitation from the dayside closed field lines despite small fluxes. The red-line signature of the diffuse aurora helps us compare diffuse and discrete aurora evolution.

The red-line emission started to brighten at 1540:21 UT at ~-75° MLAT, 2 min after the green-line aurora. This pattern became brighter until 1542:30 UT, and it then decayed and disappeared at 1544:40 UT. This structure was located near the dawnside edge of the imager FOV and it is hard to examine whether it is discrete or diffuse aurora. Unlike the green-line diffuse aurora, this pattern essentially stayed at the initial position.

At 1542:30 UT, ~4 min after the emergence of the green-line diffuse aurora, another auroral brightening emerged near the center of the imager FOV. At this time, the brightening was at the similar magnetic longitude but higher magnetic latitude than the diffuse aurora in the same snapshot. Then it became brighter, larger and also extended and propagated duskward. It had sharp edges and small structures, indicating that this is a discrete aurora. As shown in Figure 1D, the discrete aurora had a similar dawn-dusk extent to that of the diffuse aurora, and propagated duskward at a slightly smaller rate. We did not examine equatorial mapping because the T01 model predicts that these structures are on open field lines.

Since discrete aurora is formed by electron precipitation that carries upward FACs, FACs are expected to be associated with the discrete aurora brightening. In fact, Paper 1 shows that dayside high-latitude ground magnetometers in the northern hemisphere had transient oscillations in the ULF frequency range. We calculated equivalent horizontal current distributions using magnetometer data available through SuperMAG. The method is described in Nishimura et al. [*2016*]. Here, generators of FACs are in the magnetosphere, and the two ionospheres are the load of the electrical circuit. The sense of FACs relative to the ground is thus the same in the both hemispheres (upward FACs are present on roughly the same field line in the two hemispheres). This allows us to compare the upward FAC in the northern hemisphere and aurora in the southern hemisphere.

Figure 6 shows equivalent horizontal currents in green lines and positions of the magnetometers are shown by green dots. To see the current variances better, we subtracted the currents at 1534 UT as background, when both solar wind and aurora were steady. At ~1536 UT, the upward FAC shown in the pre-noon sector and another current shear at post noon, indicate the existence of a region-2 (R2)-sense current. This timing corresponds to the initiation of the diffuse aurora. Then this R2 sense current disappeared quickly. At 1541 UT, a pair of the R1-sense current emerged near the noon. The counterclockwise horizontal current vortex, which corresponds to an upward FAC of this R1-sense current, became more intense and propagated duskward (~13 MLT at 1541 UT in Figure 6c; ~15 MLT at 1546 UT in Figure 6d). The positions of this current vortex are consistent with the positions of the high-latitude red-line auroral pattern shown in Figure 5, supporting the inference that the discrete auroral location and propagation display evolution of upward FACs due to the foreshock transient-magnetopause interaction. Here, the consistency of the initiation times of the horizontal currents and the aurora brightenings observed by the ASI at the south-pole also suggests that the interhemispheric comparison is reasonable.

## 6 Discussion

In this event, the green-line diffuse aurora was detected a few minutes before the red-line discrete aurora, and was located equatorward of the discrete aurora, both of which propagated duskward. Ground-based magnetometers also show a pair of R2-sense FACs during the diffuse aurora followed by a clear duskward-moving anticlockwise horizontal current vortex (upward

FAC) associated with the discrete aurora. These signatures are analogous to auroral and ionospheric current responses during dynamic pressure pulses except that the scale size is smaller: Nishimura et al. [*2016*] observed a pair of R2 and then R1-sense currents associated with a discrete aurora brightening after a diffuse aurora brightening, and identified this as the response to primary (PI) and main (MI) impulses of sudden commencement (SC) events. This similarity indicates that magnetosphere and ionosphere responses to the foreshock transient possibly follow the same physics as during SCs in a smaller scale size, namely magnetosphere compression and enhanced R2 and R1-sense FACs. Also, the association between the red-line aurora and ground magnetic field is analogous to that during TCVs [*Vorobjev et al.*, 1993]. TCVs have been shown as a response of upstream foreshock transients [*Sitar et al.*, 1996; *Murr et al.*, 2003], consistent with what we found in this study.

This interpretation is also consistent with the magnetopause compression observed by all 5 satellites in the magnetosphere. However, the magnetopause crossing only lasted for a few minutes and the solar wind dynamic pressure measured by THEMIS-B and C did not change before and after the foreshock transient. Thus the measured magnetospheric and ionospheric responses are not due to an interplanetary shock but due to the foreshock transient.

The time delay between the in-situ and ionosphere signals would arise from a finite response time of wave-particle interaction to the magnetosphere compression, M-I communication time, and the imager time resolution. *Chi et al*. [2006] showed that the dayside signal propagation time is about a minute. The time delay we found is roughly a full bounce period of Alfven waves between the two hemispheres, as possibly a time scale for the (R2-sense) current system to develop.

Figure 7 shows a schematic illustration of magnetospheric and auroral responses to the foreshock transient on June 25, 2008. The foreshock transient locally compresses the magnetopause, and launches ULF waves in the magnetosphere and FACs. Energetic electron precipitation associated with upward FACs and enhanced wave-particle interaction drives localized discrete/diffuse aurora brightening. The auroral location and size (~4.3 RE as maximum) indicate the MLT and finite azimuthal extent of the foreshock transient interacting with the magnetopause. The duskward propagation of aurora reflects the duskward propagation of the foreshock transient as it swept through the magnetosheath while impacting the magnetopause.

## 7 Conclusion

This paper provides the analysis of the foreshock transient evolution/propagation in 2D perspective on June 25, 2008, based on the observation by the THEMIS and GOES satellites, ASI at South Pole station, and magnetometers in Greenland and Canada. Although it is generally difficult to identify 2-D evolution of interaction between a foreshock transient and magnetospheric responses due to limited observation points, the present study shows that the foreshock transient is geo-effective and their interaction with the magnetosphere can be imaged using dayside aurora more precisely in 2-D perspective than by other means of measurements.

1. A foreshock transient was observed by THEMIS B and C to propagate toward the Earth. The transient is probably a foreshock cavity or possibly a proto-HFA, which was sandwiched within 2 tangential discontinuities, and swept through the magnetosheath from dawnside to duskside. A magnetopause crossing and ULF waves were detected in the subsequent ~10 minutes.

2. Diffuse aurora brightened soon after the foreshock transient observation. The mapped pattern of diffuse aurora brightening in the equatorial plane shows a width of ~4.3 Re in GSM-Y direction, which is slightly larger but similar to the typical size of foreshock transients. The average velocity of the mapped pattern is ~100.0 km/s in GSM-Y direction, similar to the motion of the upstream foreshock transient. The diffuse aurora also brightened in coincidence with the occurrence of R2-sense FAC.
3. Discrete aurora brightened soon after diffuse aurora brightened. The discrete aurora brightening also shows a localized signature and propagated duskward. The equivalent current pattern shows that the discrete aurora brightening is associated with the upward FAC portion of the R1-sense FAC.

The substantial geophysical impacts, including magnetospheric compression, aurora brightening and FACs, which are driven by foreshock transients, are analogous to those during SCs except at a smaller scale size. A statistical perspective study of this phenomena is warranted for the future.

# 8 Acknowledgements

This work was supported by NASA grants NNX15AI62G, NSF grants PLR-1341359 and AGS-1451911, and AFOSR FA9550-15-1-0179 and FA9559-16-1-0364. The THEMIS mission is supported by NASA contract NAS5-02099. The South Pole imager has been supported by a cooperative agreement between NSF and the National Institute of Polar Research, Japan. THEMIS and SuperMAG data were obtained from http://themis.ssl.berkeley.edu as daily CDF files, http://supermag.jhuapl.edu/ as daily ASCII files. We gratefully acknowledge support from

NSF award ANT-0638587. HZ is partially supported by NSFC 41628402 and NSF AGS-1352669.

**Table Caption**

**Table 1** Timings of the signatures in the upstream and downstream. The first column shows the MLTs of 5 satellites (Geotail12, 10, THEMIS-E, D and A from top to bottom) in the magnetosphere; the second and second columns show when the first and second discontinuities contacted the bow shock at those locations, respectively; the fourth column shows the magnetoshpheric signatures observed by the 5 satellites; the last column shows when the diffuse aurora patterns arrived in those 5 locations.

| Location [MLT] | First discontinuity contacted the bow shock [UT] | Second discontinuity contacted the bow shock [UT] | Magnetospheric signatures [UT] | Diffuse aurora pattern [UT] |
|---|---|---|---|---|
| **10.3** | 1527:55 | 1531:11 | 1533:00 [G12] | 1536:00 |
| **11.3** | 1529:11 | 1533:25 | 1534:00 [G10] | 1538:20 |
| **13.2** | 1531:45 | 1537:49 | 1539:00 [THE] | 1542:40 |
| **14.0** | 1533:01 | 1539:55 | 1541:00 [THD] | 1544:30 |
| **15.5** | 1536:00 | 1544:41 | 1544:00 [THA] | 1548:00 |

**Figure Caption**

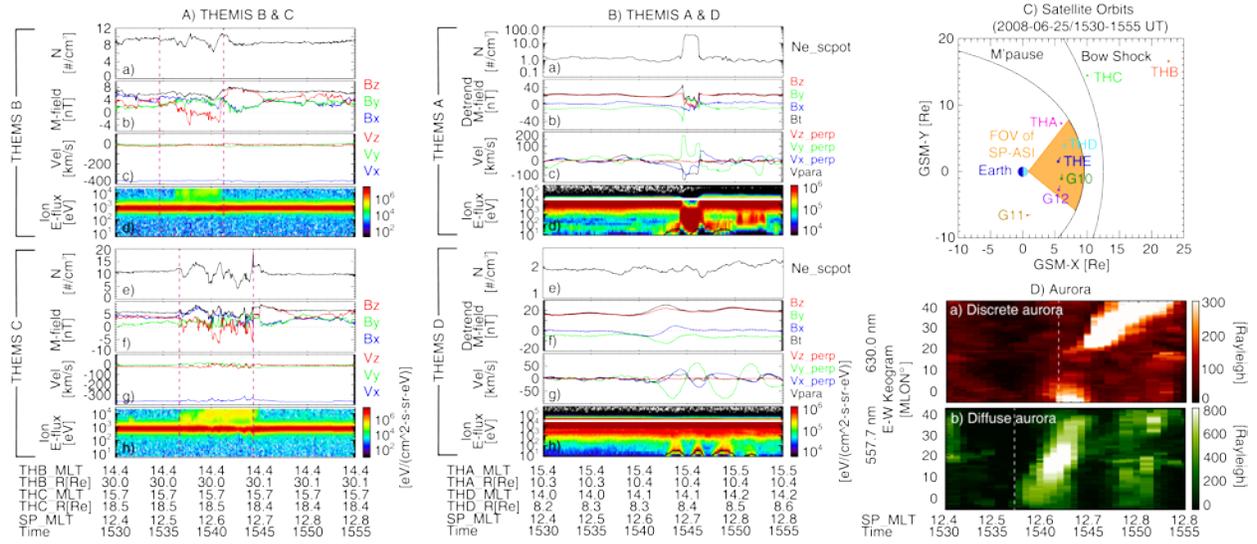

**Figure 1** Panels Aa-Ad show the plasma density, magnetic field, flow velocity and ion energy flux observed by THEMIS B during 1530 UT through 1555 UT. Panels Ae-Ah show the same format as Panels Aa-Ad, except for THEMIS C. The magenta dashed lines in Panel A present the IMF discontinuities. Panels Ba-Bd show plasma density, detrended magnetic field, flow velocity and ion energy flux observed by THEMIS-A during 1530 UT through 1555 UT. Panels Be-Bh show the same format as Panels Ba-Bd, except for THEMIS-D. Black lines in Panels Bd and Bh show the kinetic energy by the bulk flow moment. Panel C shows the orbits of THEMIS satellites and the FOV of the ASI at South Pole station during 1400 UT through 1600 UT. The positions of magnetopause and bow shock are obtained by the models described in *Shue et al.*, [1998] and *Wu et al.*, [2000], respectively. Panels Da and Db show the keograms of red-line emission and green-line emission, respectively, during 1530 UT through 1555 UT. Here, in order to show variations clearly, the background is subtracted in original snapshots (Panels 2 and 4) and E-W keograms (Panel 1D), where the background is defined as the median luminosity at each pixel within ±10 min from each observation time. White dashed lines show the initial timings of aurora brightening.

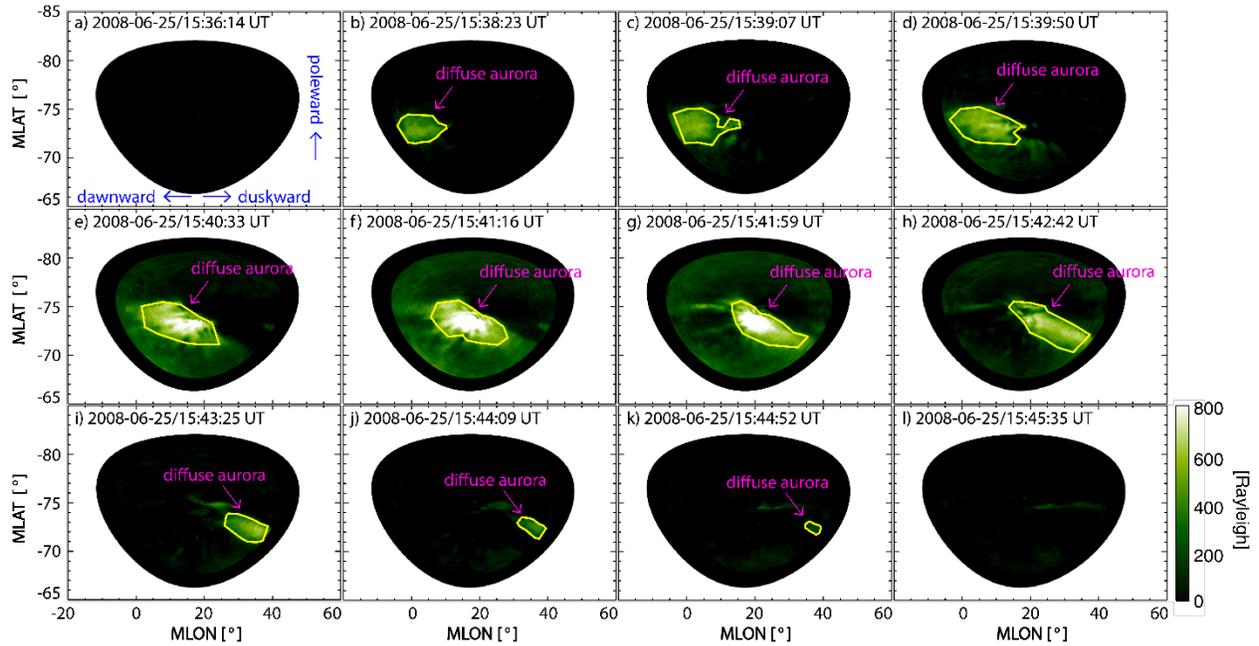

**Figure 2** Snapshots of green line (557.7 nm) emission observed by the ASI at South Pole station during 1536:14 UT through 1545:35 UT on June 25, 2008. The x-axis is magnetic longitude and the y-axis is magnetic latitude. Yellow contour lines identify patterns of diffuse aurora.

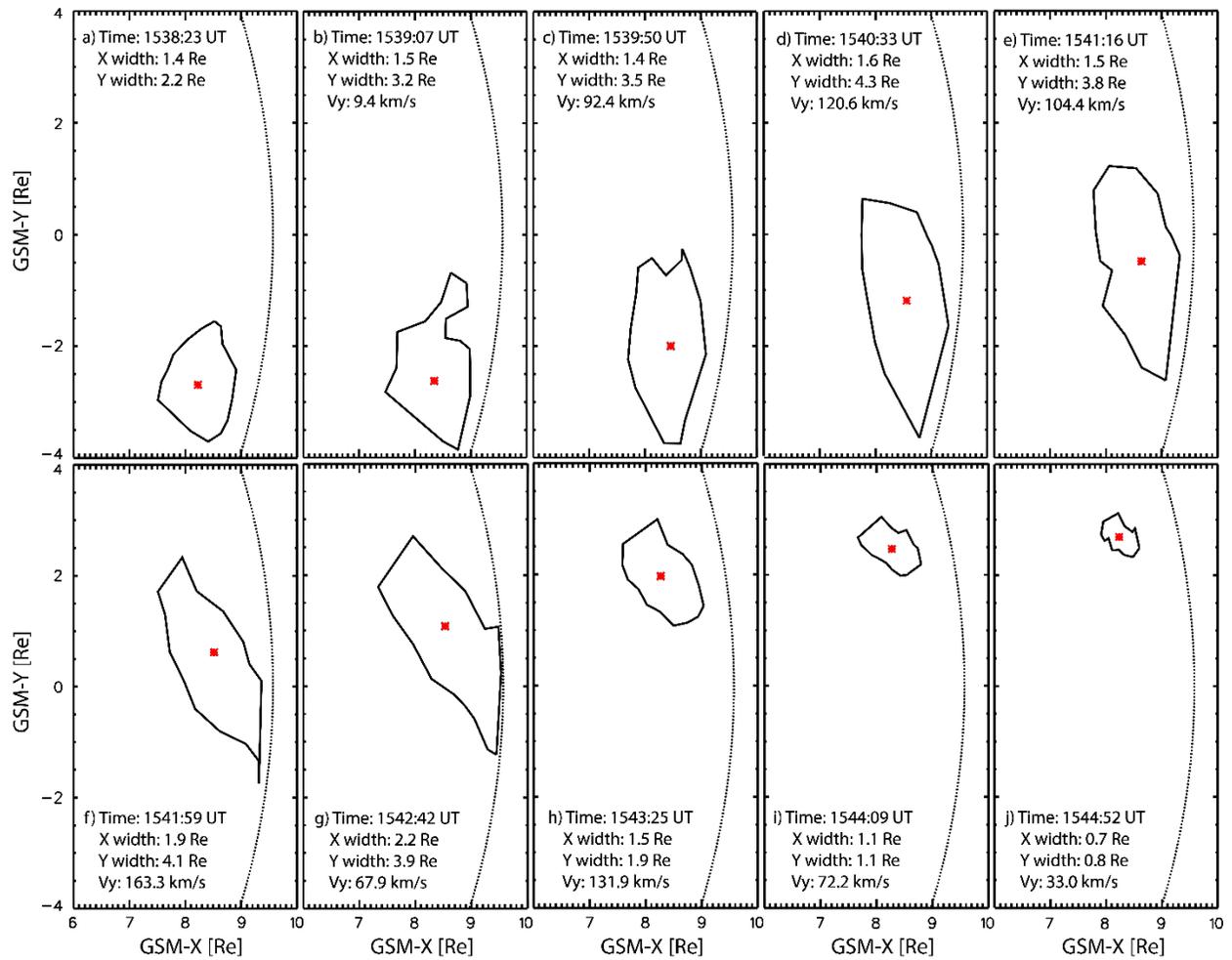

**Figure 3** Mapped patterns of diffuse aurora brightening in the equatorial plane by the T01 model during 1539:50 UT through 1544:52 UT on June 25, 2008. Red stars show mapped position of the centroid of diffuse aurora brightening patterns. The dashed curves show the position of magnetopause [*Shue et al.*, 1998].

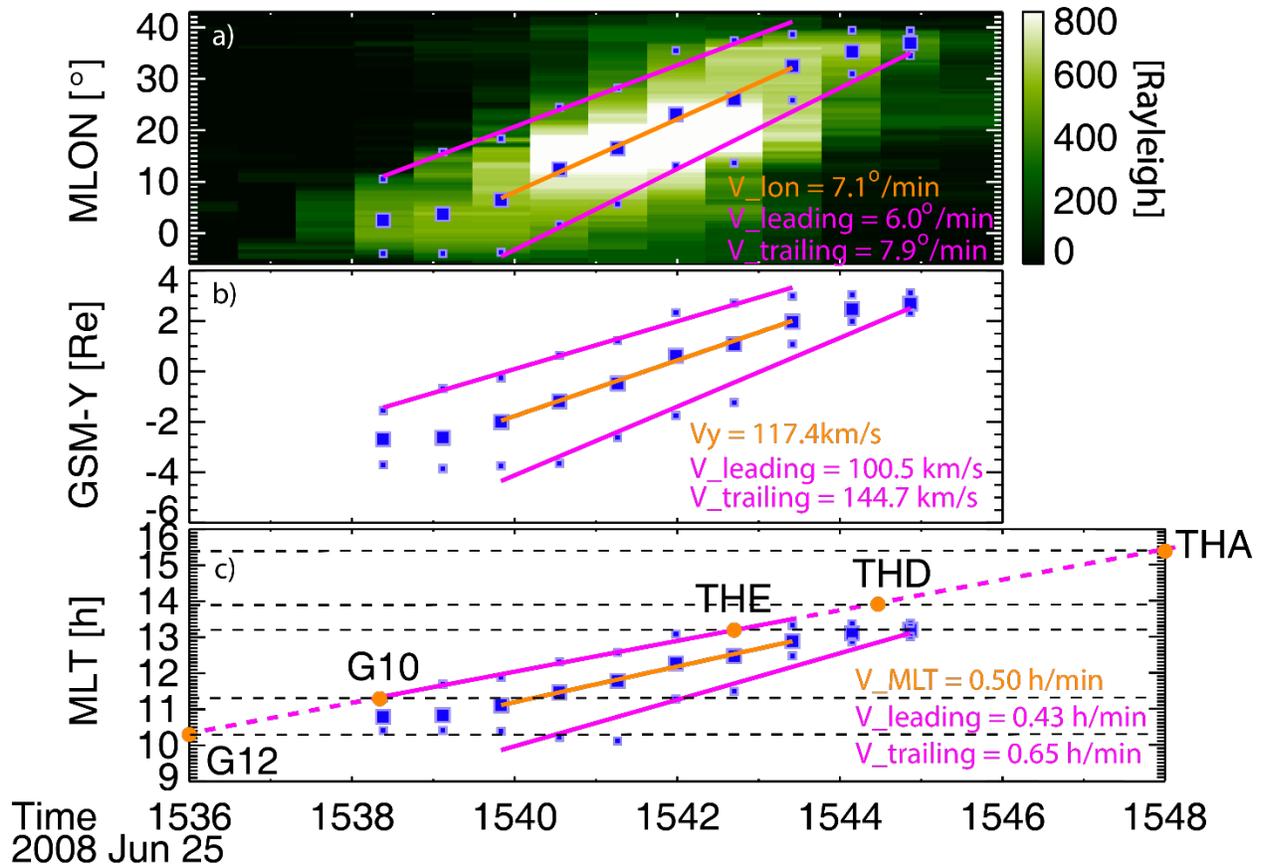

**Figure 4** Panel a shows the keogram of the green-line (557.7 nm) emission. The larger blue squares present the centroids of the patterns of diffuse aurora brightening in the ionosphere. Panels b and c show the GSM-Y and MLT positions of the mapped centroids by the T01 model to the equatorial plane. The smaller blue squares present the edges of the patterns of diffuse aurora brightening. The orange and magenta lines describe the linear fitting of those points. The orange dots show the MLT position of GOES 12, 10, THEMIS-A, D and E.

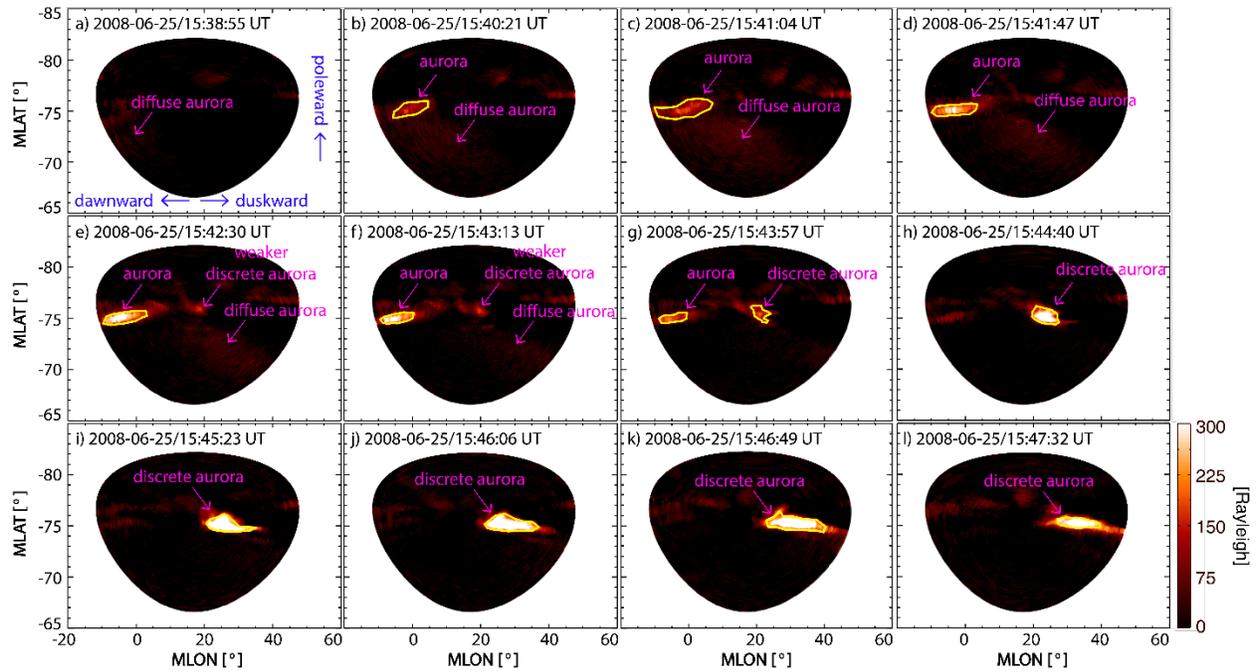

**Figure 5** Snapshots of red-line emission observed by the ASI at South Pole station during 1538:55 UT through 1547:32 UT on June 25, 2008. The x-axis is magnetic longitude and the y axis is magnetic latitudes. Yellow contour lines identify patterns of discrete/diffuse aurora.

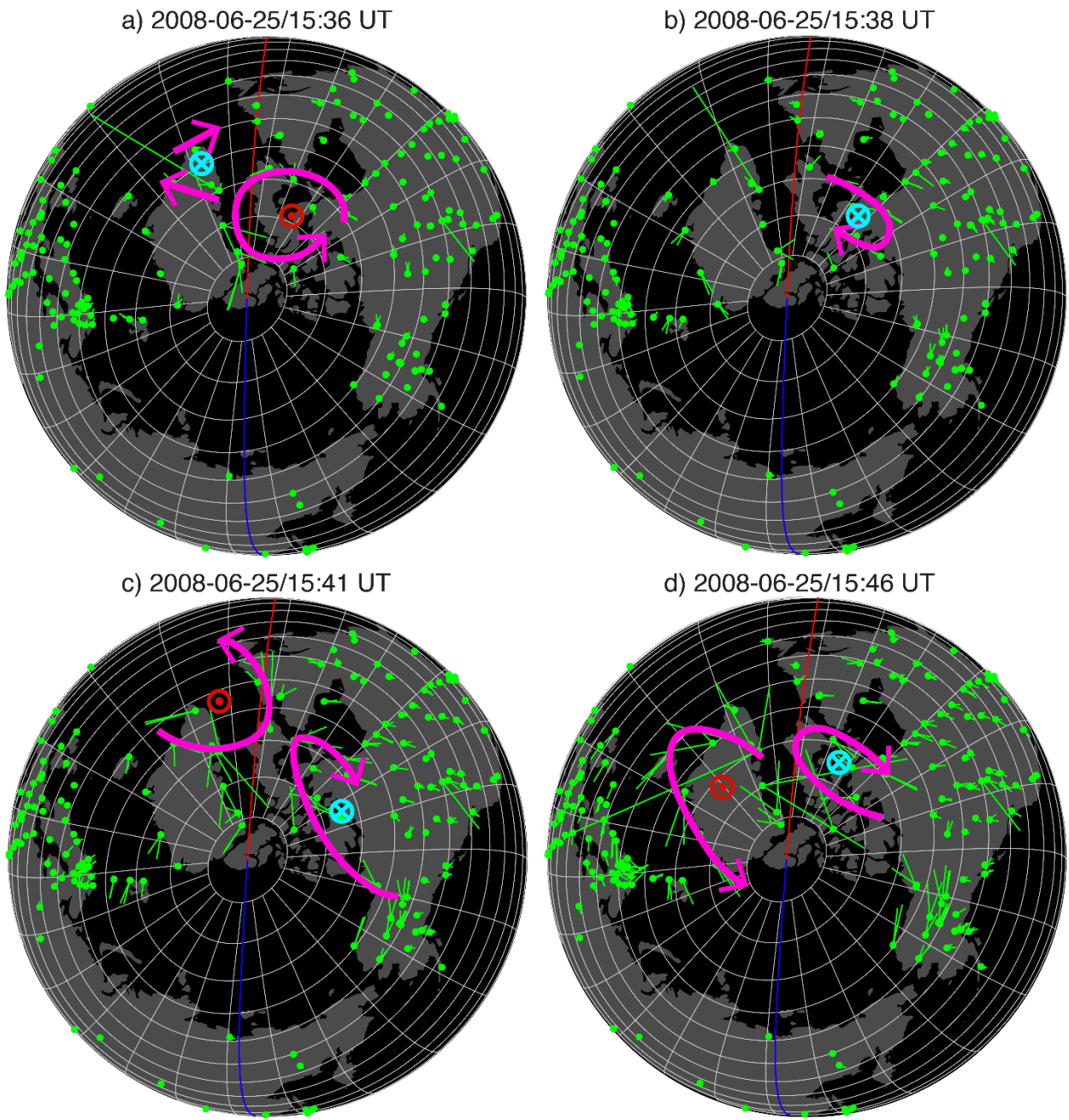

**Figure 6** Horizontal equivalent currents that are calculated from the magnetic field observed by ground-based magnetometers in the northern hemisphere during 1536 UT through 1546 UT on June 25, 2008. Green dots show the position of the magnetometer and green lines present the direction and magnitude of horizontal current. Red and blue lines indicate the noon and midnight meridians, respectively. Magenta circles schematically indicate current vortexes.

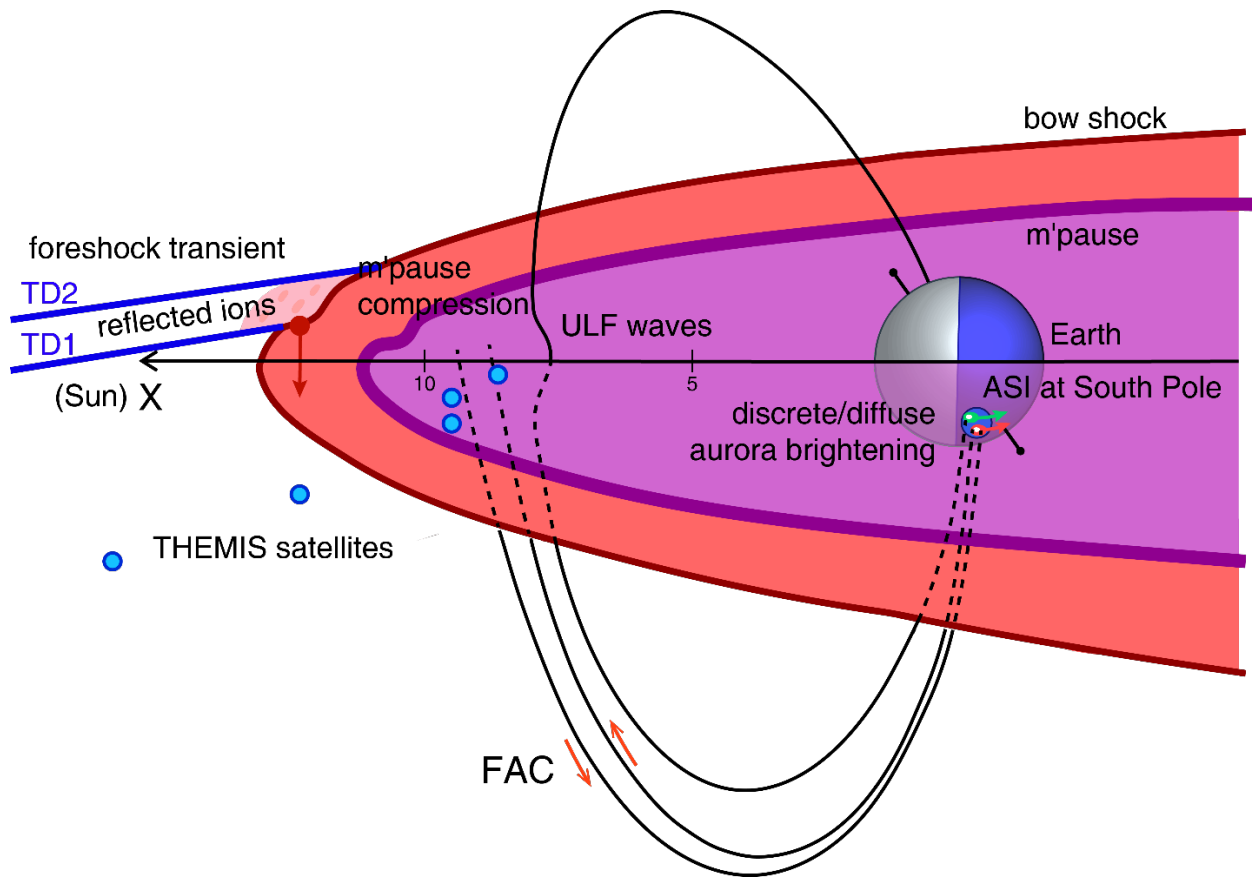

**Figure 7** Schematic illustration of magnetospheric and aurora responses to the foreshock transient on June 25, 2008.